\documentclass[%
 reprint,
 amsmath,amssymb,
 aps,
]{revtex4-1}

\usepackage{graphicx}
\usepackage{dcolumn}
\usepackage{bm}
\usepackage{hyperref}
\usepackage{epstopdf}
\usepackage{eucal}

\usepackage[sort&compress]{natbib}
\newcommand{\R}{\mathbb{R}}
\newcommand{\C}{\mathbb{C}}

\graphicspath{{Figures/}{./Figures/}}

\begin{document}


\title{Dynamic mode decomposition for multiscale nonlinear physics}

\author{Daniel Dylewsky}
 \affiliation{Department of Physics, University of Washington, Seattle, WA 98195.}
 \email{dylewsky@uw.edu}


\author{Molei Tao}
\affiliation{
School of Mathematics, Georgia Institute of Technology, Atlanta, GA 30332
}%

\author{J. Nathan Kutz}
\affiliation{%
 Department of Applied Mathematics, University of Washington, Seattle, WA 98195
}%


\date{\today}

\begin{abstract}
We present a data-driven method for separating complex, multiscale systems into their constituent time-scale components using a recursive implementation of {\em dynamic mode decomposition} (DMD). Local linear models are built from windowed subsets of the data, and dominant time scales are discovered using spectral clustering on their eigenvalues. This approach produces time series data for each identified component, which sum to a faithful reconstruction of the input signal. It differs from most other methods in the field of {\em multiresolution analysis} (MRA) in that it 1) accounts for spatial and temporal coherencies simultaneously, making it more robust to scale overlap between components, and 2) yields a closed-form expression for local dynamics at each scale, which can be used for short-term prediction of any or all components. Our technique is an extension of {\em multi-resolution dynamic mode decomposition} (mrDMD), generalized to treat a broader variety of multiscale systems and more faithfully reconstruct their isolated components. In this paper we present an overview of our algorithm and its results on two example physical systems, and briefly discuss some advantages and potential forecasting applications for the technique.

\end{abstract}

\pacs{Valid PACS appear here}
\maketitle


\section{\label{sec:intro}Introduction}
Physical systems whose dynamics evolve on a broad range of scales simultaneously (spatial or temporal) have been the subject of much study in the development of diagnostic and modeling tools. These multiscale systems are ubiquitous in physics, so there is a great deal of practical interest in methods which are accommodating to scale disparities spanning orders of magnitude. Of particular note are those systems whose behavior can be decomposed into a finite number of discrete scales, as this lends an additional structural constraint which can be exploited in modeling.  For instance, atmospheric climate data and/or simulations can be characterized by developing separate models for variations on the order of one day and one year, respectively, and then coupling them. Mathematical methods for exploiting these distinct and disparate scales can greatly simplify the problem of state estimation and forecasting.  We extend the method of {\em dynamic mode decomposition} (DMD) in order to characterize multiscale physics and their coupling dynamics, showing that such a data-driven strategy provides a viable and adaptive strategy for diagnostics and dynamical modeling. 

The task of identifying distinct multiscale temporal physics directly from data in a way that allows the signal to be decomposed into its constituent scale-separated components is a subject of ongoing investigation. Well-established methods use Fourier- and wavelet-based techniques to separate coarse-grain and fine-grain features in space or time (but generally not both at once) \citep{daubechies92, ganesan04, weniger17}. These approaches, though often useful, are purely diagnostic. They do not directly produce dynamical models from data. Moreover, their focus on exclusively temporal (or exclusively spatial) coherencies limits their utility as precursors to model discovery for state estimation and forecasting.  Regardless, such techniques form the mathematical basis of {\em multiresolution analysis} (MRA)~\citep{mallat89, kutz2013data} which provides a rigorous foundation for multiscale decomposition.

To address the dynamical limitations of MRA, researchers have put forth a number of equation-free, data-driven modeling techniques tailored to multiscale spatio-temporal systems. Indeed, there is a significant body of research focused on modeling multiscale systems and linking scales: notably the heterogeneous multiscale modeling framework, equation-free methods, and structure preserving versions known as FLAVORs \citep{kevrekidis_equation-free_2003,weinan_heterognous_2003,weinan_principles_2011,tao10}. Additional work has focused on testing for the presence of multiscale dynamics so that analyzing and simulating multiscale systems is more computationally efficient \citep{froyland_computational_2014,froyland_trajectory-free_2016}. Many of the same issues that make modeling multiscale systems difficult can also present challenges for model discovery and system identification~\citep{Champion2019}. This motivates the development of specialized methods for performing model discovery on problems with multiple time scales, taking into account the unique properties of multiscale systems.  A purely data-driven approach was recently introduced by Kutz. et. al.\citep{kutz16} which recursively applies DMD to build closed-form linear models to approximate dynamics at all scales simultaneously. DMD was first proposed as a decompositional technique for complex fluid flows \citep{schmid10, rowley09, schmid11}, but it has since been adopted more widely as a method for finite approximation of the Koopman operator in a large variety of data sets \citep{tu14}. DMD produces a linearized model for a (generically nonlinear) data set. It can be thought of as a best-fit approximation of a signal generated by a linear combination of static spatial modes whose time-varying weights follow complex exponential trajectories of oscillation, growth, or decay. The technique proposed by Kutz et. al., dubbed {\em multi-resolution Dynamic Mode Decomposition} (mrDMD), builds on MRA wavelet techniques by recursively subdividing the data set to access different regimes of the time-frequency domain. The length of the window over which DMD is applied is repeatedly halved, and the most salient components of each iteration are interpreted as a simplified local model for the dynamics at that scale. 

Decomposition of a multiscale signal can also be cast as a Blind Source Separation (BSS) problem by treating each time scale as an independent source contributing to the composite signal being measured. Typical methods for BSS include Principal Component Analysis (PCA) and Independent Component Analysis (ICA). A comparison of these methods to DMD is presented in \citep{kutz2016dmd}. When the source signals occupy fairly narrow frequency bands, as is assumed to be the case in this paper, DMD is shown to drastically outperform the other techniques. For this reason it is an obvious candidate for the decompositional method used in the sliding window framework we present here.

This paper aims to extend and generalize the mrDMD algorithm. The essential insight of mrDMD is the sensitivity of results to duration of the input signal: given a time series containing dynamics on widely-varying time scales, the eigenfrequencies obtained by DMD could reflect any of these time scales depending on the duration and resolution of sampling. The window lengths tested in mrDMD are limited to some base time span and power-of-two subdivisions thereof. This can be problematic in systems whose multiscale frequency content does not follow that pattern---the ability of DMD to robustly identify a persistent component at a particular time scale turns out to be fairly sensitive to window size. The simple halving scheme could easily fail to resolve a component whose characteristic time scale falls between those given by powers of two. We solve this problem by 1) implementing a protocol using sliding, overlapping windows on the data set to generate spectral bands of DMD eigenvalues, and 2) developing a diagnostic to use the narrowness of these bands to tune the window size for optimal resolution of a particular scale component.  The method we propose here effectively identifies and isolates the constituent time scale components of two test systems. In addition to providing diagnostic information on the frequency content of a signal, it produces 1) faithful reconstructions of each constituent component with minimal cross-pollution between them, 2) closed-form expressions for these reconstructions which can be used for low-cost forecasting at any time scale, and 3) statistics on the parameters of windowed DMD models, whose distributions can be sampled for stochastic ensemble forecasting. 

The method we present bears some similarity to the Frequency Map Analysis (FMA) technique often used in analysis of time series generated by nonlinear dynamics \citep{laskar03}. Both seek to identify dominant frequencies in a multivariate signal and fit the data to a linear combination of sinusoids at these frequencies. However, FMA does this with a static basis of spatial modes (corresponding to the canonical coordinates of Hamiltonian mechanics) whereas our approach allows spatial modes to vary over time using a sliding-window framework. This makes it more versatile in its ability to reconstruct a wide variety of input signals. Moreover, FMA typically restricts its analysis to real-valued frequencies which produce purely sinusoidal dynamics. Our method can be similarly constrained, but in general it admits complex-valued frequencies which also allow for exponential growth or decay in its local windowed reconstructions.

The rest of the paper is outlined as follows: In Sec. \ref{sec:background} we present an overview of the theory and implementation of DMD. In Sec. \ref{sec:methods} we outline the protocol for our sliding-window scale separation technique and demonstrate it on a simple toy model. In Sec. \ref{sec:scale_sep_performance} we briefly discuss the advantages of our method over traditional temporal filtering tools. A fully-fledged recursive, many-scale example using data from a three-body planetary system is presented in Sec. \ref{sec:3body}. The paper is concluded in Sec. \ref{sec:discussion} with a discussion of theoretical context for our approach and possibilities for its future application.

\section{\label{sec:background}Background: Dynamic Mode Decomposition}
Dynamic Mode Decomposition (DMD) seeks a best-fit linear model for a time-series data set. 
Given some collection of sequential measurements $\bm{x}_j \in \R^N$ for $j = 1, \ldots, M$, DMD solves for an operator ${\bf A}$ which satisfies, to closest approximation, ${\bf x}_{j+1} \approx {\bf A} {\bf x}_j$ for all snapshots $j$. This can be computed by separating the data ${\bf X} \in \C^{N\times M}$ (where $N$ is the dimension of the measurement space and $M$ is the number of data points measured) into two sequential matrices ${\bf X}_1\in \C^{N\times (M-1)}$ and ${\bf X}_2 \in \C^{N\times (M-1)}$:
\begin{equation}
\begin{split}
{\bf X}_1 &= \left[\begin{matrix} 
| & | &  & | \\
{\bf x}_1 & {\bf x}_2 & \cdots & {\bf x}_{M-1}\\
| & | &  & |
\end{matrix}\right]\\
{\bf X}_2 &= \left[\begin{matrix} 
| & | &  & | \\
{\bf x}_2 & {\bf x}_3 & \cdots & {\bf x}_{M}\\
| & | &  & |
\end{matrix}\right]\\
\end{split}
\end{equation}
The operator ${\bf A}$ is then simply the matrix which minimizes the Frobenius norm $\left|\left|{\bf X}_2 - {\bf A}{\bf X}_1\right|\right|_F$ so that ${\bf X}_2\approx{\bf A} {\bf X}_1$. This is a straightforward computation, but can become prohibitively expensive for large state dimension $N$. Indeed, the {\em exact} DMD algorithm \citep{tu14} approximates the operator as ${\bf A}={\bf X}_2 {\bf X}_1^\dag$ where $^{\dag}$ denotes the Moore-Penrose pseudo-inverse (least-square regression). It is therefore common to first project the data into a lower-dimensional space using Singular Value Decomposition (SVD). For a more detailed overview of the method, see Tu et. al. (2014) \citep{tu14}. 

A well-known deficiency of the exact DMD approach is the adverse effect of the measurement errors (sensor noise) on its performance \citep{ashkam18,Hemati2017tcfd,Dawson2016ef}. Previous studies indicate that presence of sensor noise can negatively influence the computation of eigenvalues and they would be biased, presenting a serious problem for studies that rely upon exact DMD to distinguish between stable and unstable modes. This is primarily because exact DMD treats data sequentially rather than as a whole, and thus favors the forward time direction. Recent studies have addressed this issue and proposed several techniques to mitigate this problem through employing various forms of ensemble averaging, cross-validation, windowing, and rank reduction. However, Hemati et al. \citep{Hemati2017tcfd} showed that the resulting analysis from the aforementioned techniques are subject to systematic bias errors when the measurements are inexact due to sensor noise or other effects.

In this paper, we employ a variation on the standard DMD algorithm known as Optimized DMD, which seeks to address these shortcomings. Optimized DMD recasts the minimization problem outlined above as a task of exponential curve-fitting making use of the variable projection method. This comes at the price of convexity, but yields a decomposition which much more faithfully reconstructs the input data series. A full exposition of the Optimized DMD algorithm and its advantages is presented in Ashkam \& Kutz (2018)\citep{ashkam18}. 


\section{\label{sec:methods}Methods: Time Scale Separation using DMD}
The decomposition method introduced in this paper consists of the following steps:
\begin{enumerate}
\item A sliding-window implementation of DMD to extract a large number of (complex) frequencies $\omega_j^k$ associated with spatially-coherent dynamics in the input time series
\item A clustering algorithm to identify the most highly-represented frequencies in the population of $\left\{|\omega_j^k|\right\}$. These clusters represent the multiple time scale regimes present in the input data
\item Retroactive labeling of modal components of each windowed DMD identified in Step 1 with labelling based on the cluster assignments of their associated frequencies
\item For each distinct scale regime identified in Step 2, a separate DMD reconstruction is produced by summing over the components assigned to that cluster. These reconstructions are produced separately for each iteration of the sliding-window DMD from Step 1
\item A single global reconstruction is produced for each time scale regime by combining weighted contributions from each windowed reconstruction
\end{enumerate}
Note that Steps 1-3 are {\em offline} computations. Having carried them out on a representative data set for a given system, the clustering results can be used to label new data from the same system.

\subsection{\label{sec:methods_model}A Simple Toy Model}
To introduce this decomposition method, we make use of a simple system with nonlinear dynamics on two distinct time scales. The model is given by the equations
\begin{equation}
\begin{split}
\dot{v}_1 &= v_2\\
\dot{v}_2 &= -w_1^2v_1^3\\
\dot{w}_1 &= w_2\\
\dot{w}_2 &= -\epsilon^{-1} w_1 - \delta^{-1} w_1^3
\end{split}
\end{equation}
The parameters which set the time scale separation were assigned the values $\delta = 0.25$ and $\epsilon = 0.01$, respectively. The system is initialized at $t=0$ in the state $(v_1,v_2,w_1,w_2) = (0, 0.5, 0, 0.5)$. Taken alone, the $w$ variables (i.e. the ``fast scale'') form an undamped Duffing oscillator in which the cubic nonlinearity term can be considered a small perturbation from simple harmonic motion. The $v$ variables (representing the ``slow scale'') also take the form of a cubic oscillator (sans linear term), but with a coefficient ($w_1^2$) which is dependent on the state of the $w$ variables.

This construction separates the fast and slow dynamics for the sake of interpretability. Because no such separation is guaranteed in measurements made on a real multiscale physical system, we take the additional step of applying a random linear mixing to the above coordinates:
\begin{equation}
\begin{split}
\left[\begin{matrix} x_1\\ x_2\\ x_3\\ x_4\end{matrix}\right] &= {\bf  Q}\left[\begin{matrix} v_1\\ v_2\\ w_1\\ w_2\end{matrix}\right] 
\end{split}
\end{equation}
where ${\bf Q}$ is a randomly generated $4\times 4$ orthogonal matrix. This system is numerically integrated (using Matlab's \texttt{ode45} solver) for a duration of $48$ time units with a sampling interval of $\Delta t = 4 \times 10^{-4}$. The results are aggregated into a data matrix ${\bf X}  \in \C^{N\times M}$ where $N=4$ and $M = 120,000$. 

\subsection{\label{sec:methods_mwDMD}Sliding Window DMD}
\begin{figure}
\centering
\includegraphics[width=0.9\columnwidth]{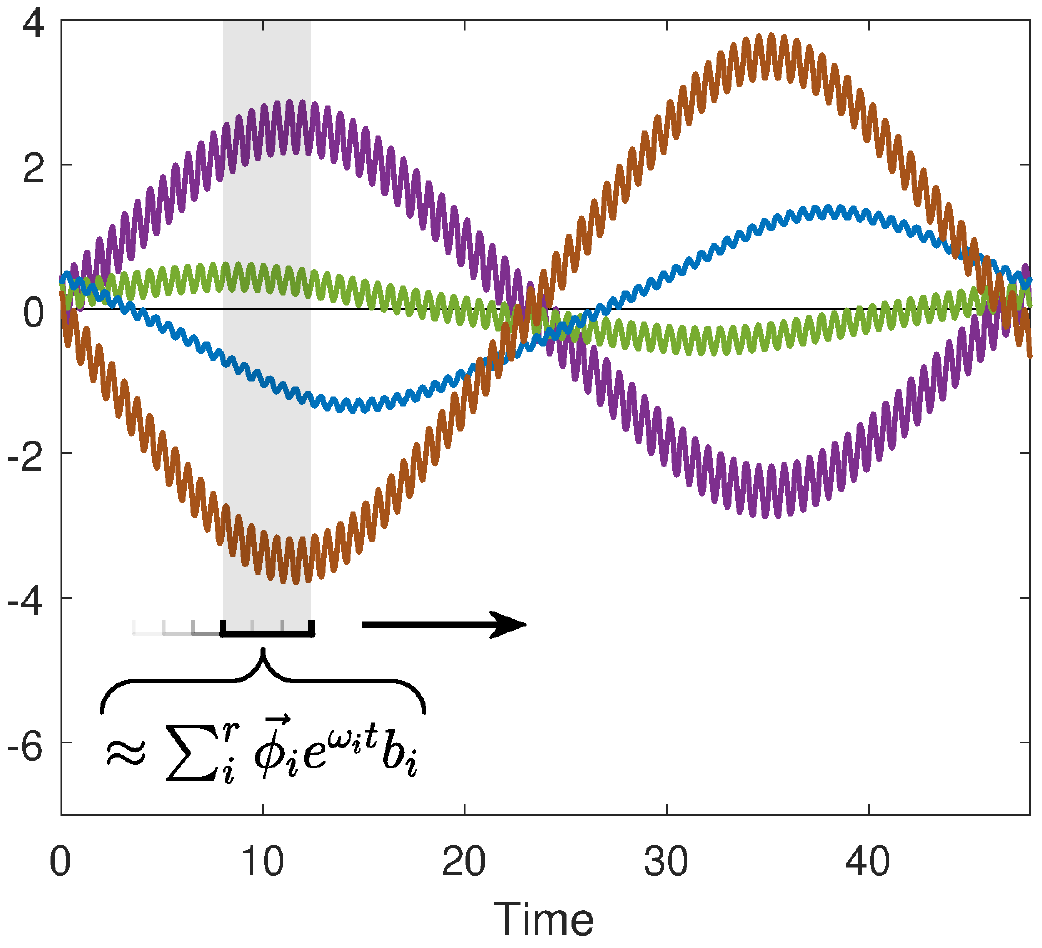}
\caption{Sliding a window filter across a longer time-series data set. For each step of the window's movement, a new DMD result is obtained. Note that the width of the window is such that it contains multiple complete oscillations of the fast scale but only a fraction of a period of the slow scale.}
\label{fig:moving_window}
\end{figure}

The sliding window approach takes advantage of DMD's sensitivity to the duration and sampling rate of the time series input it receives. Consider an $N$-variable system measured over $M$ time points. An application of DMD to the full data matrix ${\bf X}$ would  identify a frequency spectrum that would likely look quite different from that of the same algorithm applied to a subset $\bar{\bf X}  \in \C^{N\times W}$, ($W << M$). For the purposes of this investigation, the ``correct'' sample length is defined by the multiscale properties of the data: it must be long enough to capture variations on the slowest scale, but not so long as to fail to resolve the fastest scale. In the example model, the sample length $T_W$ that most cleanly separated the two distinct time scales was $T_W \approx 2~T_{\text{fast}}$ and $T_W \approx T_{\text{slow}}/20$ (The approximate periods of oscillation associated with the fast-scale and slow-scale dynamics, respectively). This is illustrated in Fig. \ref{fig:moving_window} (though the width of the window drawn has been increased slightly for visual clarity). The step size for the foreward motion of the window is chosen to be much smaller than the width of the window (about $4\%$) so that any given time point is contained by a large number of windows.

DMD approximates the dynamics contained in each window with a linear operator. The best-fit reconstruction of the windowed data can then be expressed in the standard form for solutions to a first-order linear system, using the eigendecomposition of that operator:
\begin{equation}
\begin{split}
\label{eq:dmd_recon}
\bar{\bf x}^k(t) = \sum_j^r \boldsymbol{\phi}_j^k e^{i\omega_j^k t} b_j^k + {\bf c}_k
\end{split}
\end{equation}
Here $k$ is used to index the steps of the sliding window (i.e. $\bar{\bf x}^k(t)$ is the time series contained by the $k$th window position). $j$ indexes the eigenvalues ($\omega$) and eigenvectors ($\boldsymbol{\phi}$) of the linear DMD operator (numbered from 1 to $r$ for a rank-$r$ decomposition). ${\bf c}_k$ denotes the (constant) mean of the input data, which must be reintroduced only if it was subtracted off before applying the DMD algorithm.

Note that while the form presented in Eq.~(\ref{eq:dmd_recon}) is manifestly complex, applying DMD to a real-valued input signal leads it to identify oscillatory modes in complex conjugate pairs whose imaginary components cancel each other out in the reconstruction process. In the discussion that follows, all $\bar{\bf x}^k(t)$ can be taken to be real.

\subsection{\label{sec:methods_clustering}Frequency Clustering}
The spectral results of the sliding-window DMD procedure are then clustered to discover their dominant frequency content. Concatenating the set of all $|\omega_j^k|^2$ into a single vector, cluster centroids are obtained using the $k$-medians algorithm \citep{murphy} (i.e. $k$-means using an $L_1$ distance metric to limit the influence of outliers). The choice to cluster in $|\omega_j^k|^2$ rather than $|\omega_j^k|$ has the effect of inflating the separation of higher frequencies and compressing that of lower frequencies. For this example, this has no practical effect. In the second half of this paper, however, we introduce a recursion method which uses multiple clustering iterations working sequentially from the fastest time scales to the slowest. In this case, improved differentiation between higher frequencies is an asset, and the compression of the lower frequencies is inconsequential because they can simply be dealt with on the next iteration. 

\begin{figure}
\centering
\includegraphics[width=0.9\columnwidth]{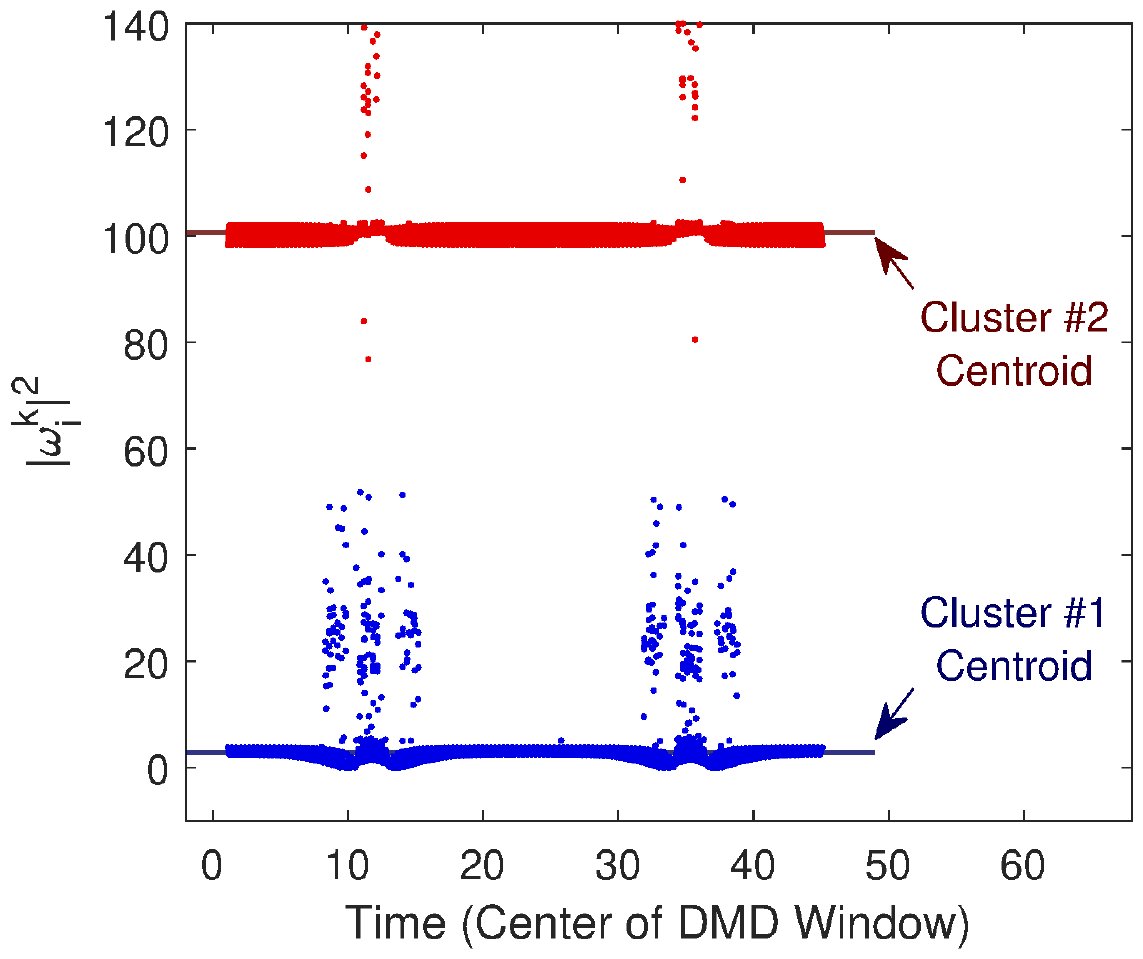}
\caption{Spectra of the (modulus squared) frequencies obtained by the sliding-window DMD procedure. Frequencies are plotted at the midpoints of the windows from which they were computed. Colors denote the cluster labels assigned to each point retroactively.}
\label{fig:om_spec}
\end{figure}

\begin{figure}
\centering
\includegraphics[width=0.9\columnwidth]{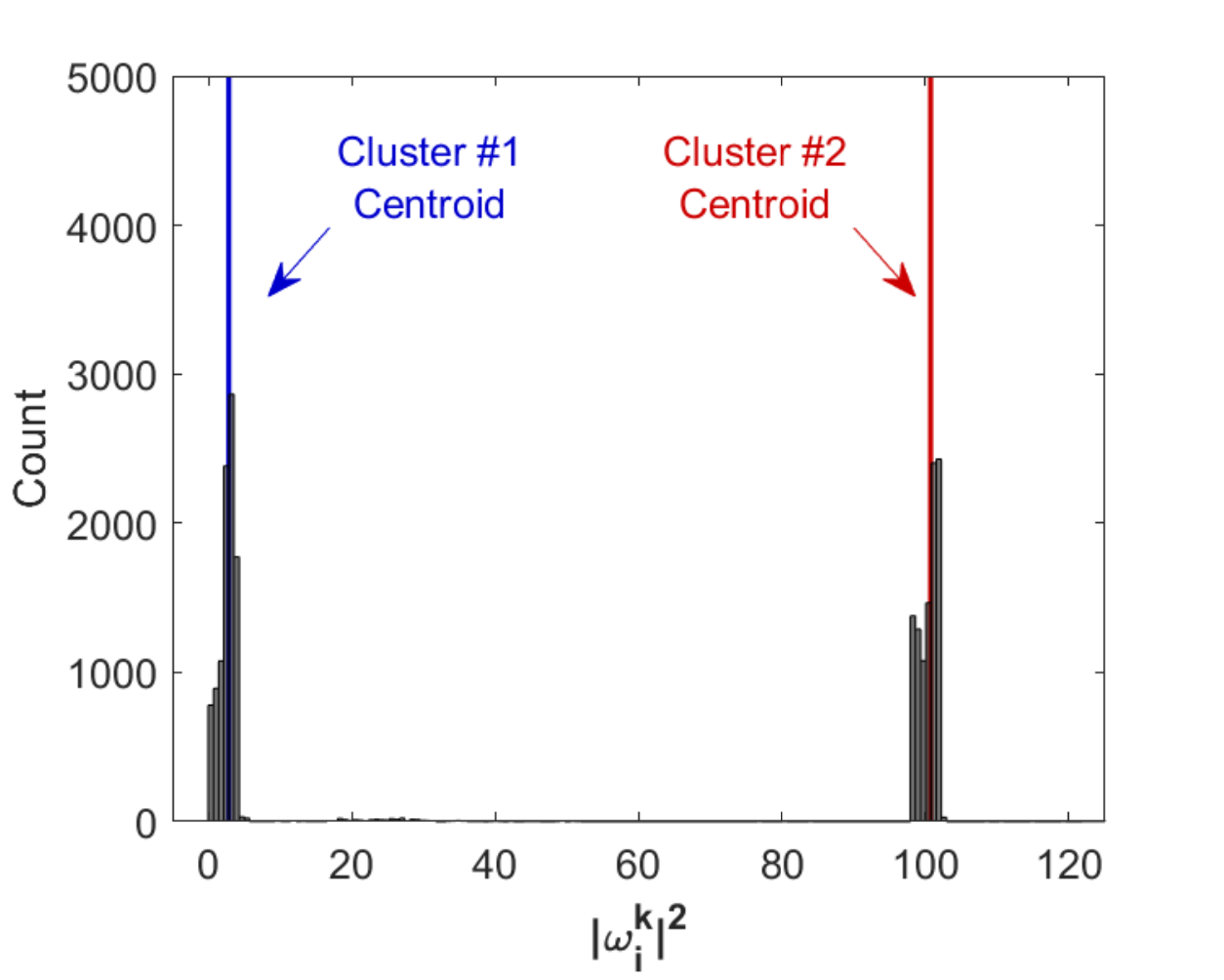}
\caption{Histogram of all $|\omega_j^k|^2$ with the $k$-medians cluster centroids overlaid in color. Note that the outliers visible in Fig. \ref{fig:om_spec} are vastly outnumbered by in-band data points}
\label{fig:om_hist}
\end{figure}

Plotted in Fig \ref{fig:om_spec} are the spectra obtained by applying DMD to each windowed subset of the sample data. The multiscale structure is immediately obvious: there are two strong bands at $|\omega|^2 \approx 100$ and $|\omega|^2 \approx 1$. Although there are a number of outliers from these dominant bands (particularly in regions where the slow-scale dynamics are relatively flat), the full set of $\left\{|\omega_j^k|^2\right\}$ is unambiguously peaked about two centroids (depicted in Fig. \ref{fig:om_hist}). Using the clustering results, we retroactively label each frequency (and, by association, each windowed DMD mode) based on its $k$-medians categorization.

A brief digression regarding clustering parameterization: $k$-medians requires that the number of clusters be supplied \textit{a priori}. In the above case the choice of $k=2$ seemed quite obvious from the band structure of Fig. \ref{fig:om_spec}. But the window size was specifically tuned to be sensitive to the fast-scale dynamics at $|\omega|^2 \approx 100$ (as described in Sec. \ref{sec:methods_mwDMD}). If there had been a third, even slower time scale (e.g. at $|\omega|^2 \approx 0.01$), it would have gone entirely unnoticed---fully subsumed by Cluster \#1---resulting in an incomplete separation of scales. This apparent failure is resolved by the recursion method which is outlined in the second half of this paper. In the meantime, we simply wish to remark that for any system with persistent dynamics on multiple, discrete time scales, each of these scale components can be resolved into a clean frequency band with an appropriate choice of DMD window length. Thus, for a given window size, choosing the number of clusters for $k$-medians can be easily accomplished by visual inspection or using the statistical cluster ennumeration method of your choice. Fully isolating all time scale components may require multiple clusterings at multiple window lengths (see Sec. \ref{sec:3body_res}), but the choice of $k$ for each of these will be independent and greatly simplified by a well-delineated band structure.

\subsection{\label{sec:methods_recon}Scale-Separated Reconstruction}
\subsubsection{\label{sec_methods_recon_global}Global Reconstruction from Windowed Results}
\begin{figure}
\centering
\includegraphics[width=0.9\columnwidth]{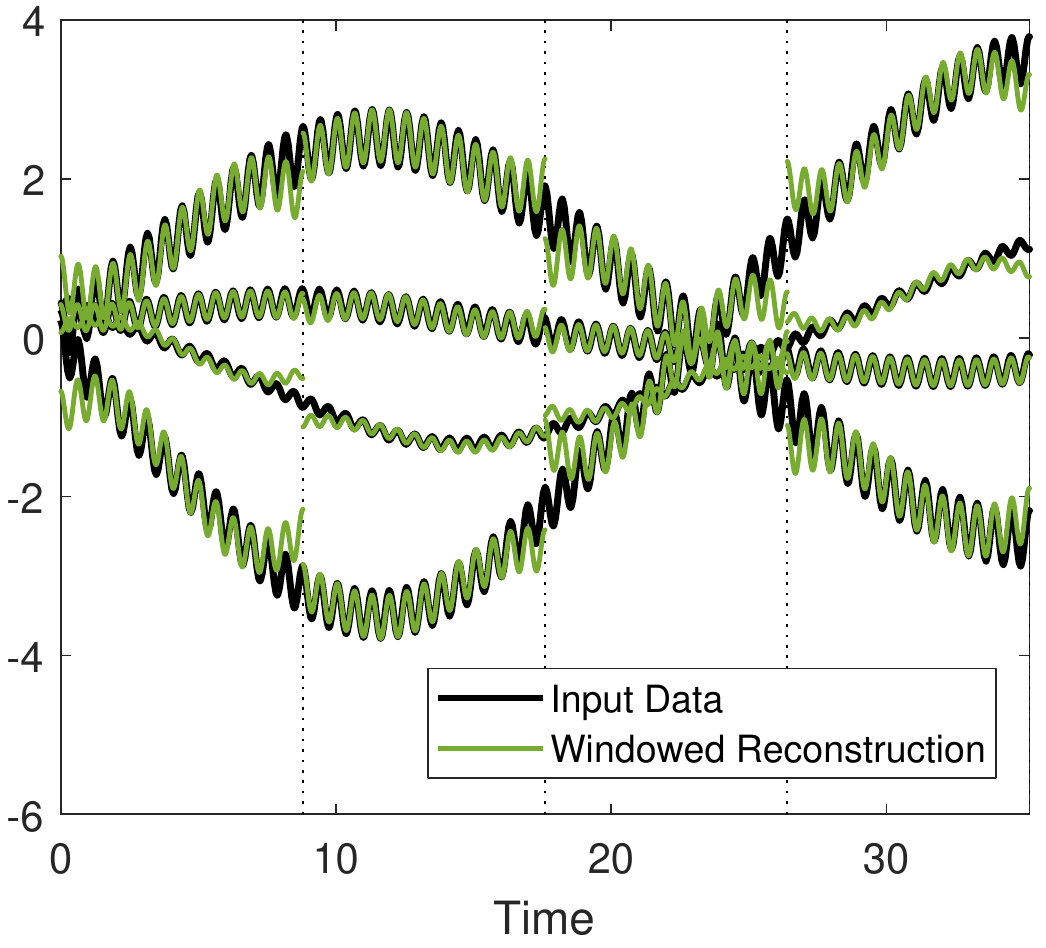}
\caption{DMD reconstructions $\bar{\bf x}^k(t)$ plotted over input data $x(t)$ for four non-overlapping windows (delineated by dotted lines)}
\label{fig:discrete_window_recon}
\end{figure}

Each windowed dynamic mode decomposition admits a linearized reconstruction of the signal, given in Eq.~ (\ref{eq:dmd_recon}). The reconstruction $\bar{\bf x}^k(t)$ (green) is overlaid on the original data $x^k(t)$ (black) in Fig. \ref{fig:discrete_window_recon}. Results are only plotted for four non-overlapping windows (demarcated by vertical dotted lines) to avoid visual clutter. 

\begin{equation}
\begin{split}
\label{eq:global_recon}
\bar{\bf x}^{\text{global}}(t) &= \frac{\sum_k e^{-(t-\mu_k)^2/\sigma^2}\bar{\bf x}^k(t)}{\sum_k e^{-(t-\mu_k)^2/\sigma^2}}
\end{split}
\end{equation}

\begin{figure}
\centering
\includegraphics[width=0.9\columnwidth]{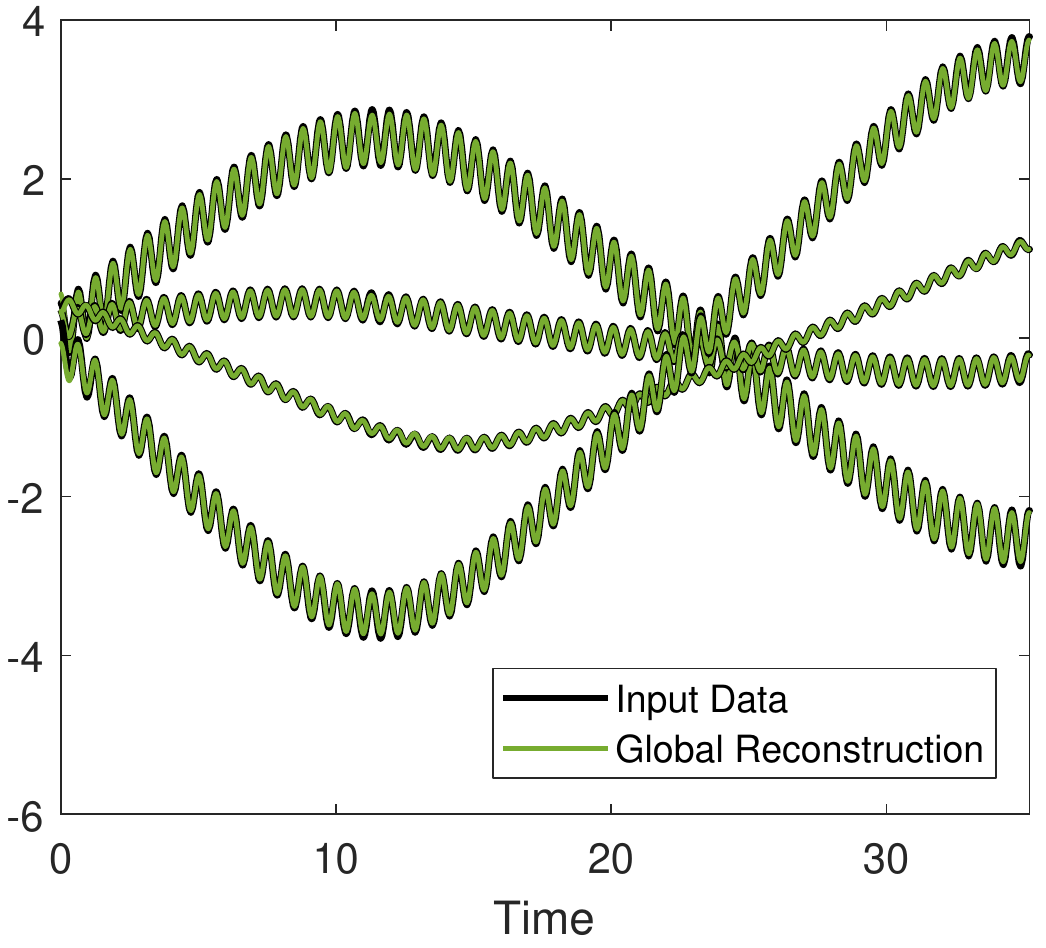}
\caption{Global DMD reconstruction $\bar{\bf x}^{\text{global}}(t)$ plotted over the input data $x(t)$}
\label{fig:global_recon}
\end{figure}

It is evident from the plot that the reconstructions tend to diverge from the true signal near the edges of each window. Converting an ensemble of windowed reconstructions into a single global reconstruction calls for a linear combination of results from all windows that contain a given point. But clearly they should not be weighted equally---to estimate the system state at $t=12$, for example, the result from the window centered on $t=12$ is more likely to be accurate than the result from a window whose boundary lies near $t=12$. To address this issue, we weight each windowed result with a Gaussian centered on the midpoint of the window $\mu_k$ and with standard deviation $\sigma$ equal to one eighth of the window's width (see Eq. \ref{eq:global_recon}).
The denominator simply acts as a normalization factor ensuring unit net contribution to every time point. The result of this method, plotted in Fig. \ref{fig:global_recon}, hews closely to the ground truth signal for the full duration of the simulation.

\subsubsection{\label{sec_methods_recon_sep}Separation of Time Scales}
\begin{figure}
\centering
\includegraphics[width=0.9\columnwidth]{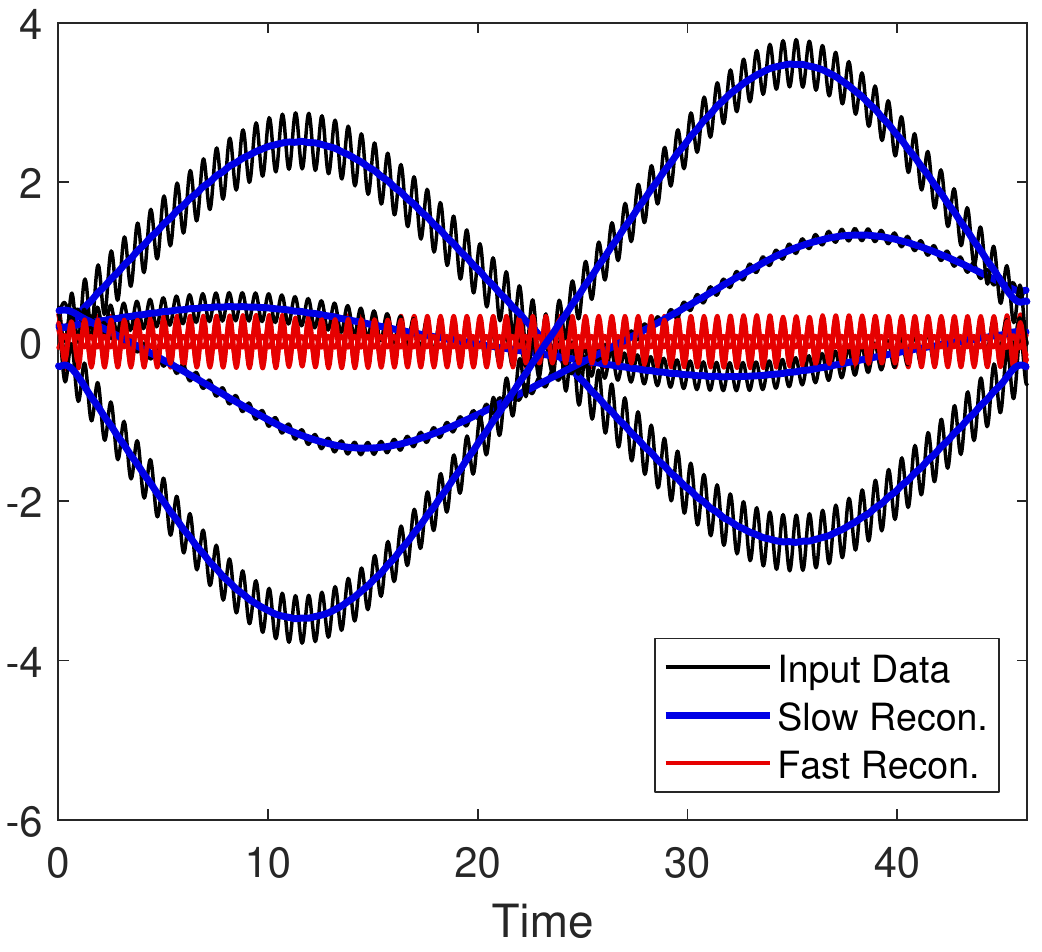}
\caption{Scale separated reconstructions $\bar{\bf x}^{\text{global}}_{\text{slow}}(t)$ (blue) and $\bar{\bf x}^{\text{global}}_{\text{fast}}(t)$ (red) plotted over the input data $x(t)$ (black)}
\label{fig:sep_recon}
\end{figure}

\begin{figure}
\centering
\includegraphics[width=0.9\columnwidth]{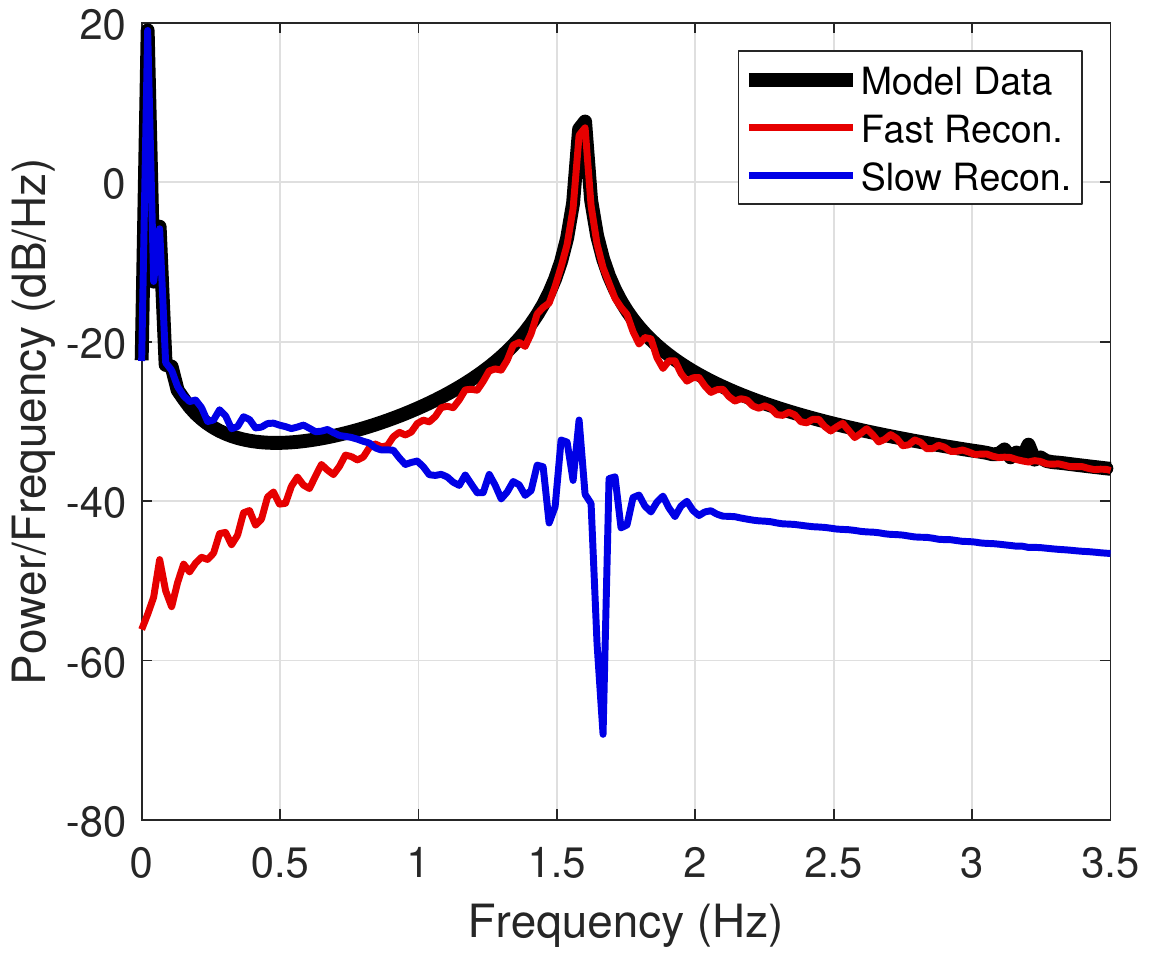}
\caption{Power spectra of the input signal (black) and fast- and slow-scale reconstructions (red and blue, respectively). The 4 variables of each signal are summed to compute frequency content}
\label{fig:sep_recon_power}
\end{figure}

Having labeled the individual modes according to the clustering results, it is straightforward to separate this summation to obtain separate reconstructions for each identified time scale:
\begin{equation}
\begin{split}
\bar{\bf x}^k_{\text{slow}}(t) &= \sum_{i\in\{\text{slow}\}}^r \boldsymbol{\phi}_j^k e^{i\omega_j^k t} b_j^k\\
\bar{\bf x}^k_{\text{fast}}(t) &= \sum_{i\in\{\text{fast}\}}^r \boldsymbol{\phi}_j^k e^{i\omega_j^k t} b_j^k\\
\end{split}
\end{equation}
In the same fashion, Eq.~(\ref{eq:global_recon}) can be separated to produce fast- and slow-scale global reconstructions (plotted in Fig. \ref{fig:sep_recon}). This result has a number of desirable properties:
\begin{itemize}
\item
Fidelity: The separated reconstructions sum to a very close approximation of the original time series (Fig. \ref{fig:global_recon})
\item
Excellent time-scale separation: There is very little mixing of frequency content between the identified regimes. Plots of the signals' power spectra (Fig. \ref{fig:sep_recon_power}) show that the separated reconstructions closely mirror the spectral content of the input signal near their respective peaks, and they contribute very little elsewhere.
\item
Spatial interpretability: Unlike other frequency filtering approaches, sliding-window DMD identifies spatial modes corresponding to dynamics of a given frequency. Concatenating the results from all windowed decompositions, we can construct time series of (complex) mode vectors which are already labeled by time scale category using the clustering results. Identifying patterns in the evolution of these modes presents a promising approach for model-building or forecasting
\item
Closed analytic form: The reconstructions $\bar{\bf x}^\text{global}_\text{slow}(t)$ and $\bar{\bf x}^\text{global}_\text{fast}(t)$ are simply weighted sums of exponentials. They therefore represent models for scale-separated variables whose values can be computed directly for arbitrary $t$ (without need for any iterative integration scheme).
\end{itemize}

These properties make this decomposition method a powerful tool for data-driven analyses of systems with multi-scale dynamics, with potential for application towards a variety of modeling and forecasting tasks.

\section{\label{sec:scale_sep_performance}Scale Separation Performance}
Given the task of separating out time-scale regimes from a multiscale signal, one standard and well-known approach is Fourier filtering. Peaks in the power spectrum could be used to identify the constituent frequencies, and each component could then be isolated using an appropriately-designed bandpass filter. This method differs from the one presented in this paper in that the former identifies only temporal coherencies in the signal, whereas the latter incorporates spatial coherencies as well. We here present a brief example of a case in which the sliding-window DMD technique outperforms Fourier filtering.

Two separate signals with different characteristic time scales are generated using two simple models:
\begin{equation}
  \begin{split}
    &\textbf{FitzHugh-Nagumo}\\[.1 in]
    \dot{v} &= v - \frac{1}{3}v^3-w+0.65\\
    \dot{w} &= \frac{1}{\tau_1}\left(v + 0.7 - 0.8w\right)
  \end{split}
\quad \quad \quad
  \begin{split}
    &\textbf{Unforced Duffing}\\[.1 in]
    \dot{p} &= q\\
    \dot{q} &= -\frac{1}{\tau_2} \left(p + p^3\right)
  \end{split}
\end{equation}
Characteristic time scales are set to $\tau_1 = 2$ and $\tau_2 = 0.2$, a factor of 10 apart. The FitzHugh-Nagumo model, used as a simple model for biological neuron dynamics, spikes sharply at intervals determined by its characteristic time scale. The Duffing model, on the other hand, is a simple nonlinear oscillator whose dynamics resemble a distorted sinusoid. Therefore, despite the disparity between $\tau_1$ and $\tau_2$, the ``slow'' component periodically acquires a rate of change comparable to that of the ``fast'' component. A combined signal $x$ is generated from these by a randomized linear mapping into $\mathbb{R}^4$:
\begin{equation}
\begin{split}
{\bf x} = {\bf  A} \cdot \left[\begin{matrix} v \\ p \end{matrix}\right]
\end{split}
\end{equation}
where ${\bf A}$ is a $4\times 2$ Gaussian-random orthonormal matrix. 

\begin{figure}
\centering
\includegraphics[width=0.9\columnwidth]{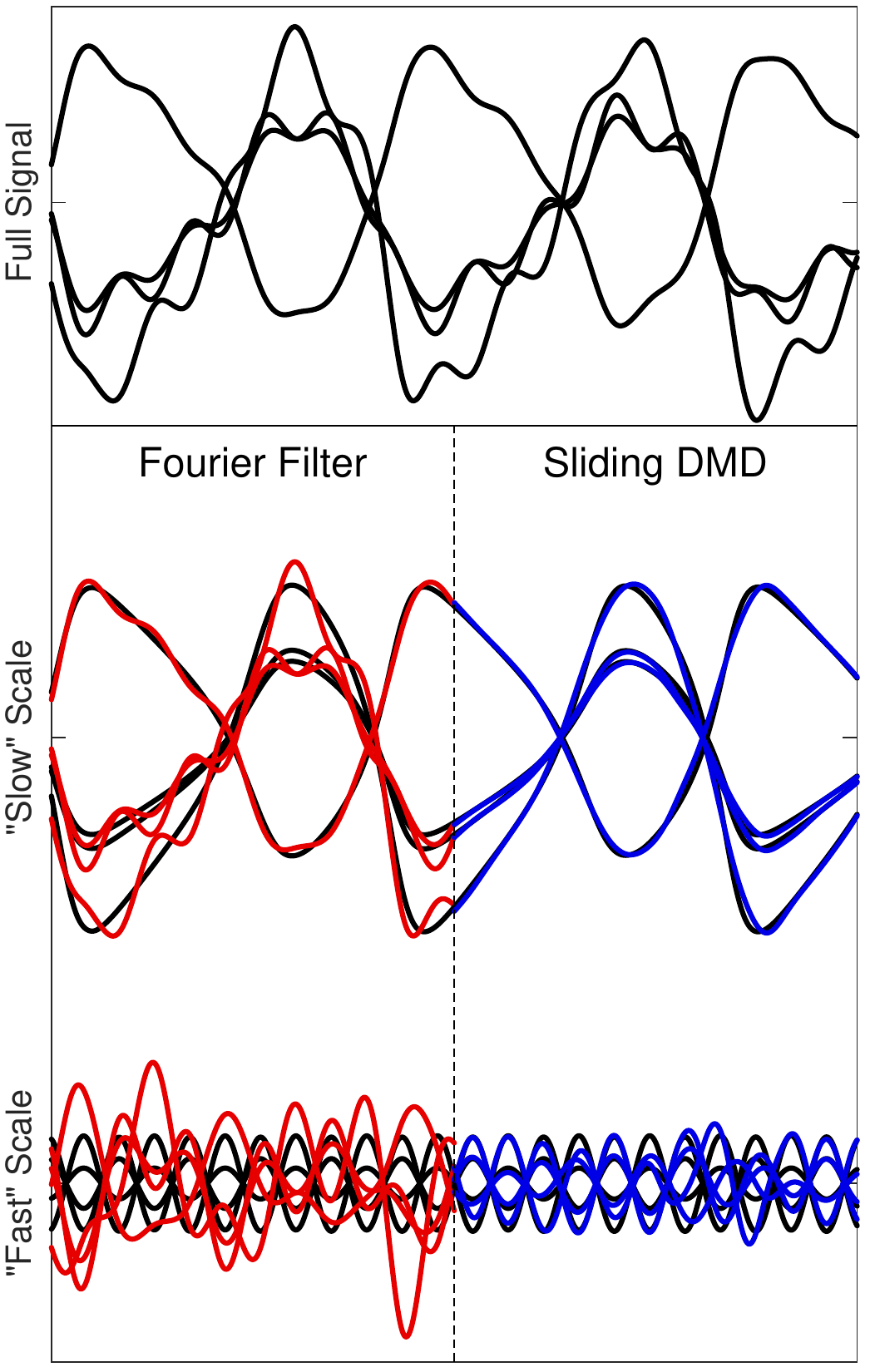}
\caption{Comparison of two scale separation techniques. The measurement signal (top) is constructed from two components, generated by the FitzHugh-Nagumo (center) and Duffing (bottom) models. The ground-truth signal separation is plotted in black, with the results of the two data-driven methods overlaid in color. The sliding-window DMD approach is much more successful in recovering the true components}
\label{fig:separation_method_compare}
\end{figure}

Signal separation is carried out using a simple Fourier filtering approach and the sliding-window DMD method. For the Fourier processing, we use Matlab's built-in low- and high-pass filter functions with passband frequencies of 0.04 Hz and 0.15 Hz, respectively. Results are plotted in Fig. \ref{fig:separation_method_compare}. 

Note that while the sliding-window DMD approach clearly performs better, neither method's reconstruction conforms perfectly to the ground truth (plotted underneath in black). Disambiguating truly overlapping scales without error is a highly nontrivial problem, beyond the scope of this paper. We present this result as evidence that sliding-window DMD is at least superior to purely-temporal methods in the case of a problem with nearly-overlapping scales, e.g. closely-spaced frequencies or nonlinear oscillations with spiking behavior.

It is also worth commenting that the data requirements for this method (i.e. duration and frequency of sampling) are at most only slightly greater than those for a Fourier-based decomposition. From an information theoretic perspective, DMD is subject to the same sampling rate restriction that applies to discrete-time Fourier analysis, i.e. the Nyquist criterion that sampling frequency must be greater than twice the highest frequency present in the signal. The lower bound on sampling duration is less strictly defined, but qualitatively it should of course be long enough to capture dynamical evolution of the lowest-frequency content of the signal. Our method introduces the additional requirement that there be enough distinct positions for the sliding window to obtain a sufficient set of frequency points for clustering. Windows can overlap with one another though, so this would (at most) perhaps double the requisite sampling duration relative to that of Fourier decomposition.

\section{\label{sec:3body}Application: A Three-Body Planetary System}
\subsection{\label{sec:3body_model} Multiscale Properties of Nearly Keplerian Orbits}
In this section we present an application of our decomposition technique to a real physical system with multiscale properties. We consider the case of three bodies interacting gravitationally in bounded orbits, with relative masses comparable to those of Jupiter, Saturn, and the Sun. Because the Sun is larger than the planets by several orders of magnitude, the system resembles two fairly stable elliptical orbits which interact weakly with one another. This suggests the presence of at least three well separated time scales in the dynamics: two ``fast'' frequencies corresponding to the planetary orbits, and one ``slow'' regime capturing the evolution of the orbits over much longer durations (which may itself have a multiscale makeup).

\subsection{\label{sec:3body_res} Recursive Application and Results}
\begin{figure}
\centering
\includegraphics[width=0.9\columnwidth]{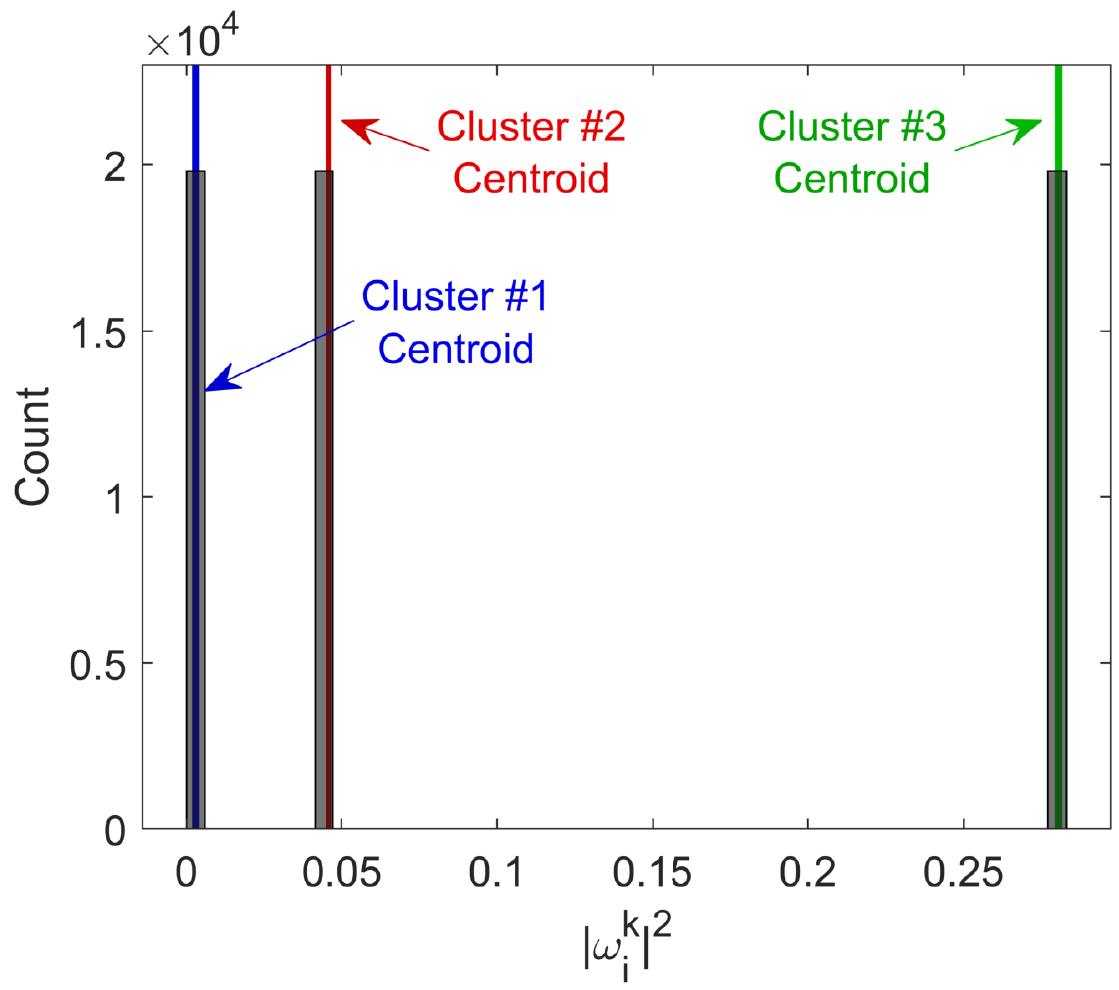}
\caption{Histogram of frequency results of sliding window DMD on the three-body planetary system. $k$-medians cluster centroids ($k=3$) are overlaid in color}
\label{fig:om_hist_3body}
\end{figure}

Data was generated for the three-body planetary system using a 4th-order symplectic integrator in Cartesian coordinates over a time span of 1,000,000 years. Applying the same sliding-window DMD procedure outlined in the previous section (window size $\sim 600$ years), the frequency content very cleanly separated into the three expected regimes (Fig. \ref{fig:om_hist_3body}). 

\begin{figure}
\centering
\includegraphics[width=0.9\columnwidth]{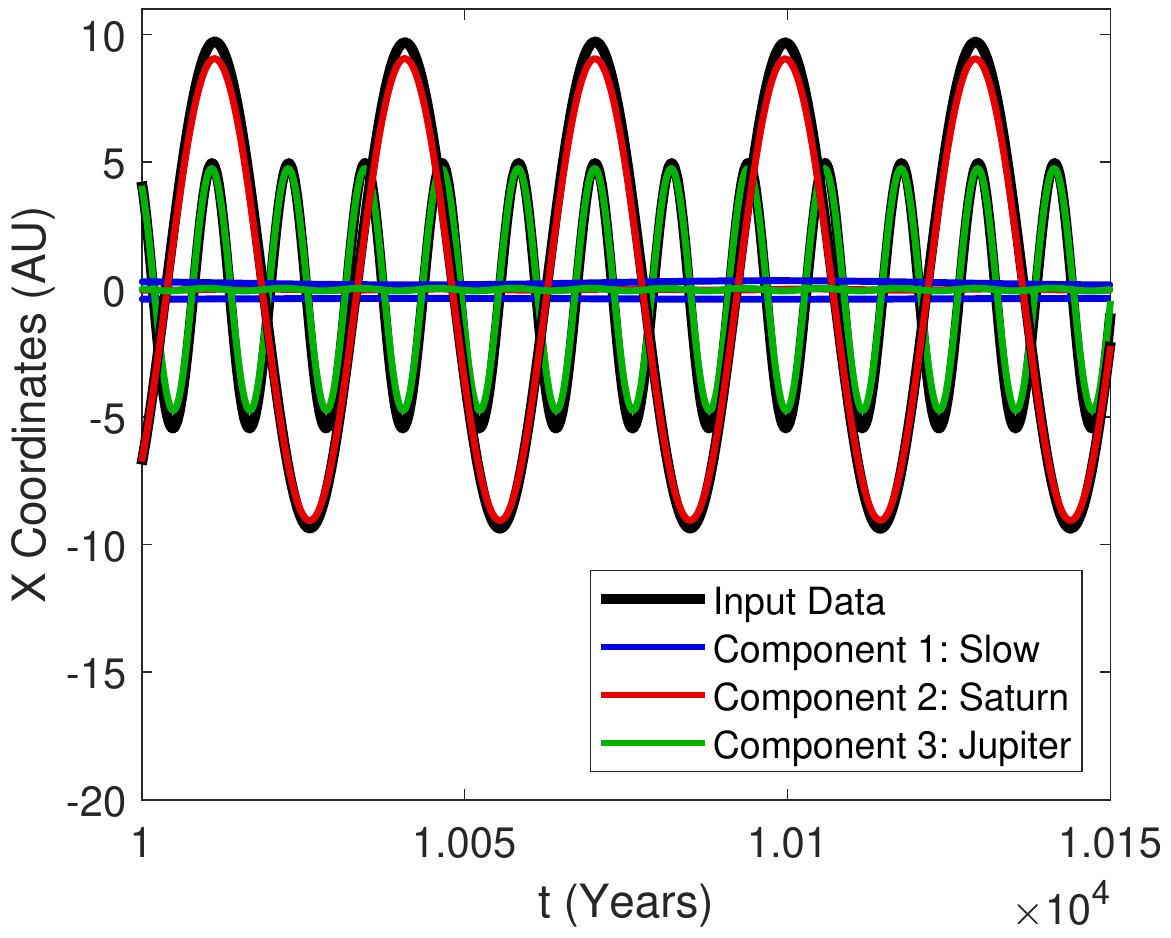}
\caption{Scale-separated reconstructions of three components (color) overlaid on the Cartesian input data (black). Note that for the short domain plotted (150 years), the slow-scale component in blue looks like a constant}
\label{fig:sep_recon_3body}
\end{figure}

\begin{figure}[htb!]
\centering
\includegraphics[width=0.9\columnwidth]{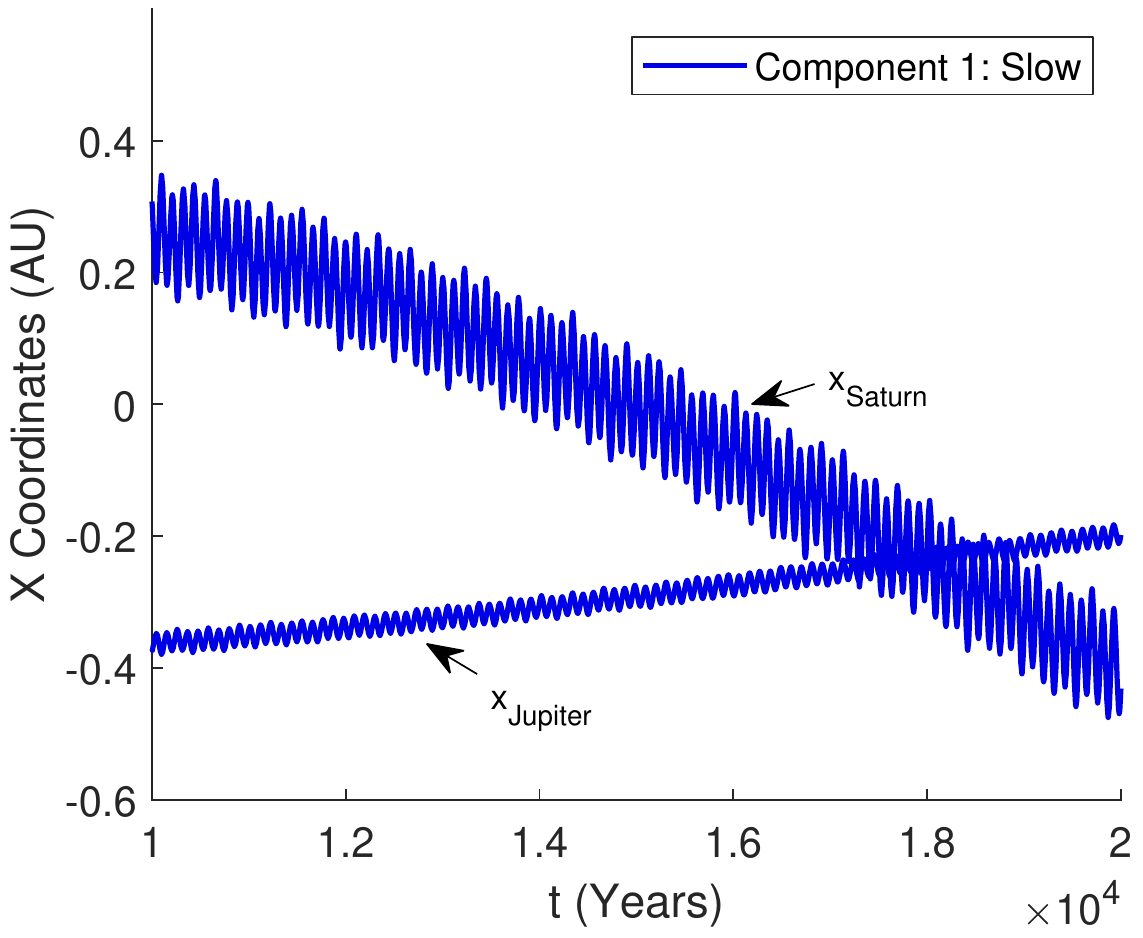}
\caption{The isolated slow-scale component obtained by the first pass of sliding-window DMD on the three-body planetary system. Plotted in blue as Component 1 in Fig. \ref{fig:sep_recon_3body}, the full scope of its dynamics is revealed over this longer time domain. The multiscale behavior evident within this single component motivates our recursive approach to scale separation}
\label{fig:sep_recon_3body_slow}
\end{figure}

Here we observe a key limitation of the sliding-window DMD approach as it has been presented thus far. The technique is sensitive to the chosen window duration, and data spanning 600 years simply does not contain sufficient information to characterize processes taking place over many millennia. Dynamics unfolding on a scale of 10,000 years would be indistinguishable from those unfolding over 100,000 years: both would just appear as a constant-valued background. While the window size used here does an excellent job of separating out the orbital frequencies of the two planets, it relegates everything taking place on time scales longer than those to a single ``slow'' regime. Zooming out on Fig. \ref{fig:sep_recon_3body} to see the evolution of this component, it is evident that it itself constitutes a rich multiscale signal with nontrivially complex dynamics (Fig. \ref{fig:sep_recon_3body_slow}).

\begin{figure}[htb!]
\centering
\includegraphics[width=0.9\columnwidth]{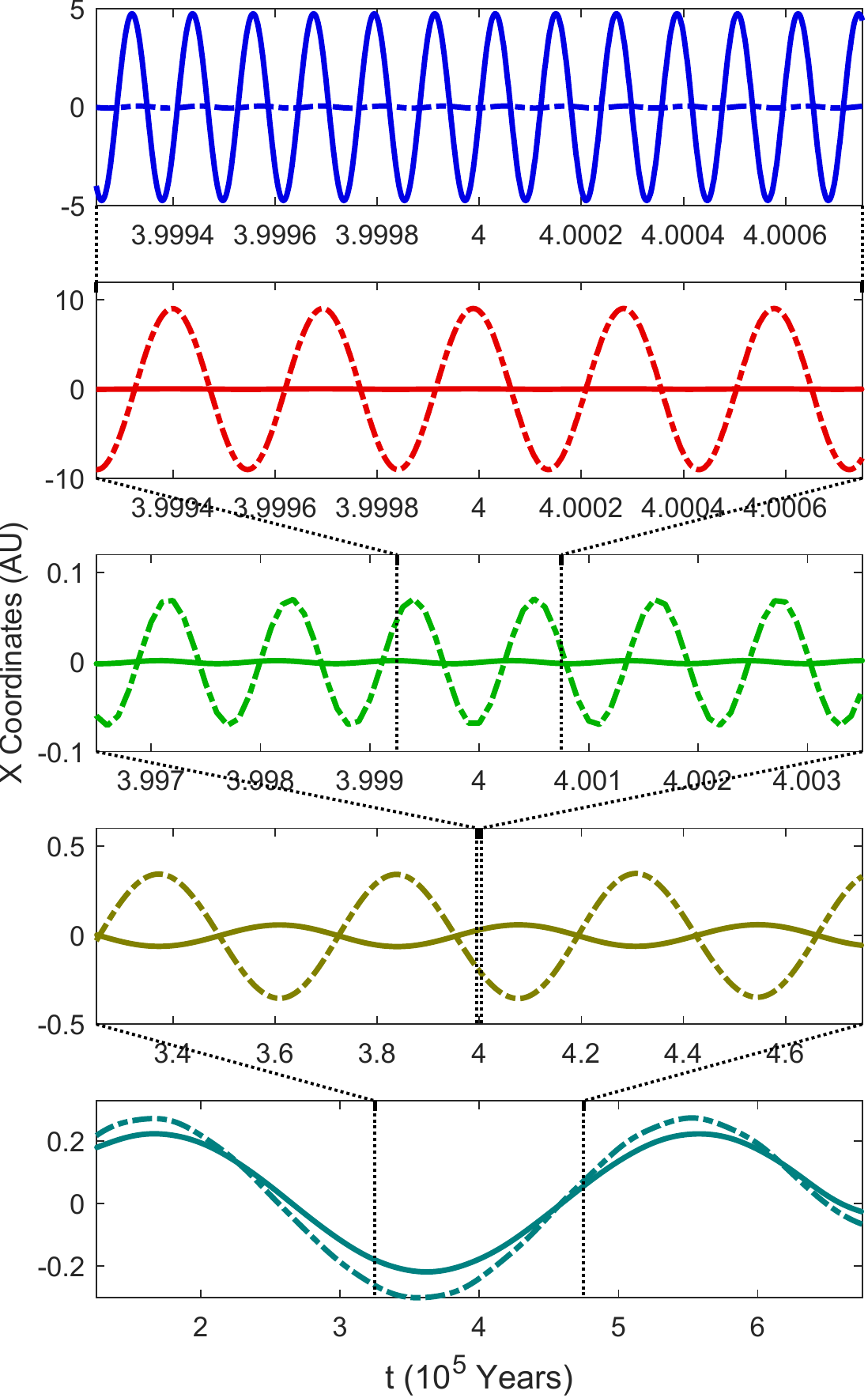}
\caption{Scale-separated reconstructions obtained from 4 recursive applications of the sliding-window DMD procedure to the three-body planetary system. Dotted lines are used to indicate the relative time scales between each successive recursion. Only the $x$ coordinates are plotted. Solid lines denote $x_{\text{Jupiter}}$ and dot-dashed lines denote $x_{\text{Saturn}}$.}
\label{fig:sep_recon_3body_recursive}
\end{figure}

\begin{figure}[htb!]
\centering
\includegraphics[width=0.9\columnwidth]{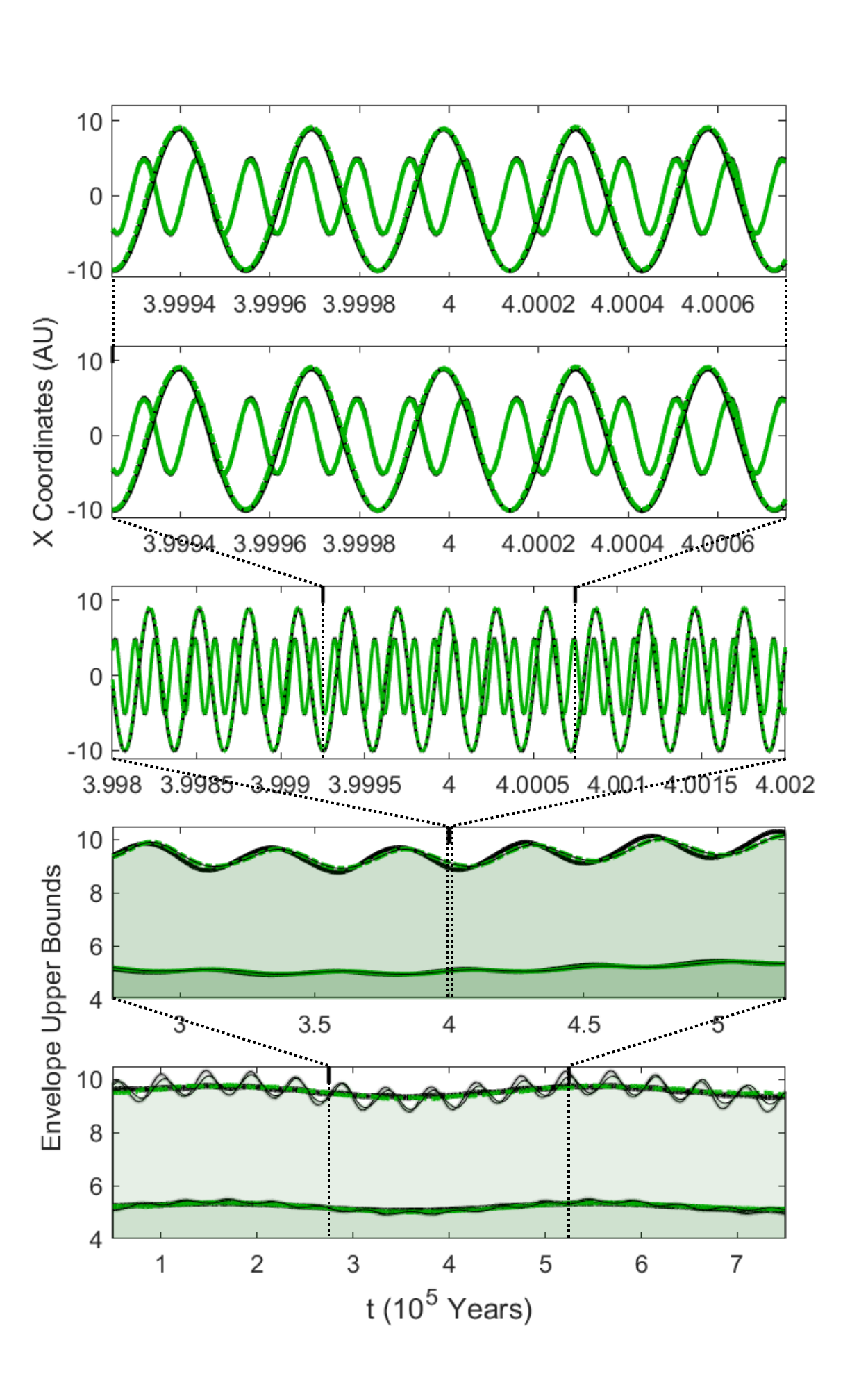}
\caption{The full reconstruction to the three-body planetary system obtained by summing the scale-separated components ($x_{\text{Jupiter}}$ in solid green, $x_{\text{Saturn}}$ in dot-dashed green) plotted over the original simulation data (black). Each plot here contains the same result, visualized over progressively longer timespans. In the last two plots, only the upper-bound envelope has been plotted for ease of visualization. The final plot shows moving averages to show the conformity even over the longest time scales}
\label{fig:full_recon_3body_recursive}
\end{figure}

To better characterize these dynamics, we recursively re-apply the sliding-window DMD approach to identify and isolate signals present in this slow component at different time scales. The methodology is identical to that of the first iteration, but now uses a window of length $\sim 4,600$ years. Repeating this process with successively longer windows, we obtain a decomposition of the original data into 5 distinct time scale components. These components are plotted separately in Fig. \ref{fig:sep_recon_3body_recursive}. The full reconstruction obtained by summing them is plotted against the input data in Fig. \ref{fig:full_recon_3body_recursive}. It successfully captures the true dynamics across all time scales.

\subsection{Physical Interpretation of Results}

\begin{figure}[htb!]
\centering
\includegraphics[width=0.9\columnwidth]{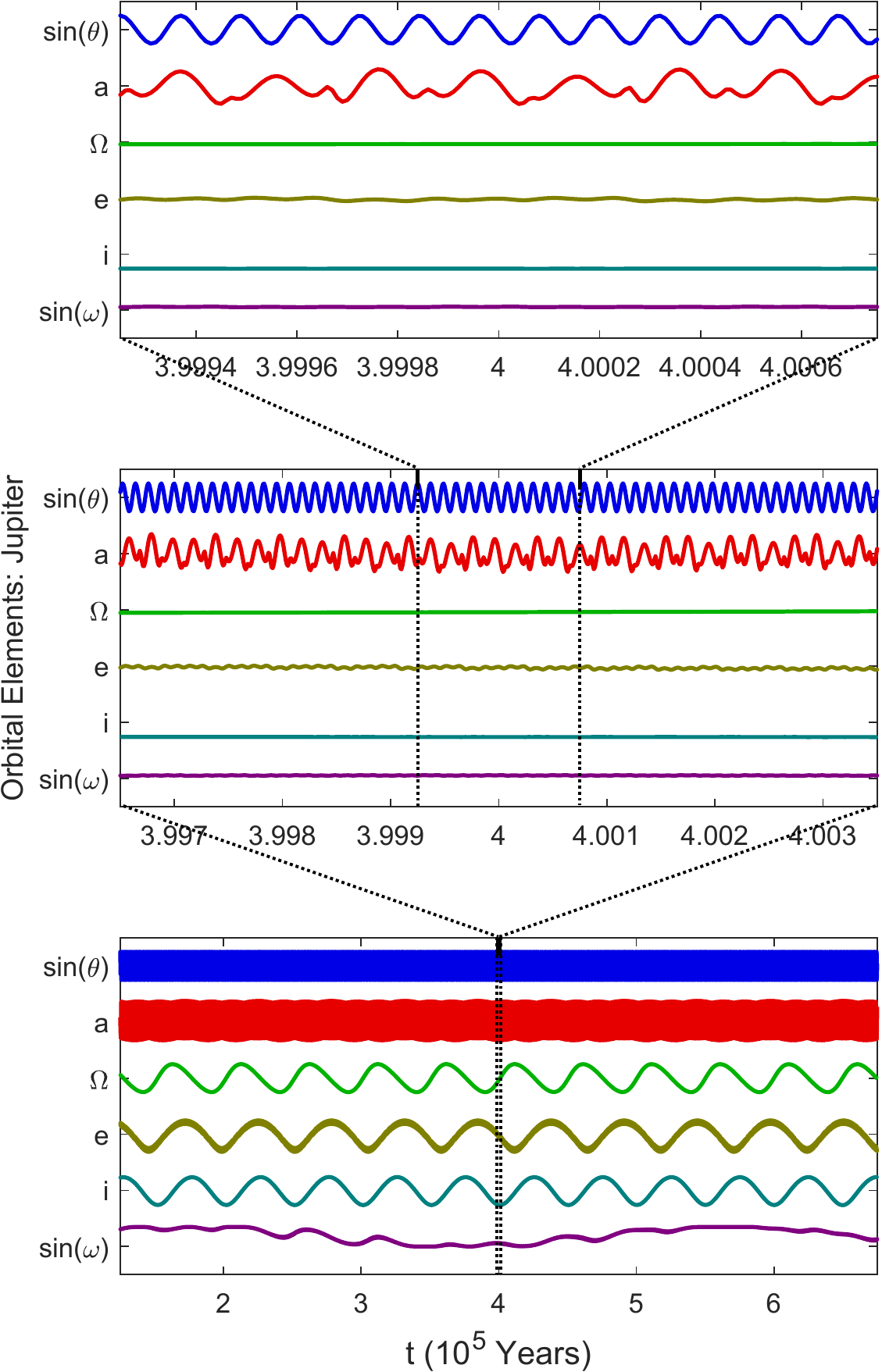}
\caption{Keplerian Orbital Elements for the orbit of Jupiter, computed from the Cartesian simulation data of the three-body planetary system. The three plots show the same data over three distinct time scales}
\label{fig:orbital_elements}
\end{figure}

We chose this example in part because it is a well-studied system with a known set of scale-separated parameters: the six Keplerian Orbital Elements provide the minimum information necessary to unambiguously define a (two-body) orbit. For each planet, we compute these quantities over the duration of the simulation: eccentricity ($e$), semimajor axis ($a$), inclination ($i$), longitude of ascending node ($\Omega$), argument of periapsis ($\omega$), and true anomaly ($\theta$). The multiscale properties of the planetary orbits can be observed by plotting these elements over different time scales. Fig. \ref{fig:orbital_elements} suggests three distinct scale regimes: $\theta$ and $a$ vary at a time scale corresponding to the planetary year, $\Omega$, $e$ and $i$ oscillate at some precessional frequency with a period of about 53,000 years, and the outer envelope of $\sin(\omega)$ has a period of over 300,000 years. 

While the DMD components we have identified (Fig. \ref{fig:sep_recon_3body_recursive}) do not all correspond precisely to the dominant frequencies of these elements, they mostly fall neatly into the same three regimes. Like $\theta$, components 1 and 2 have periods corresponding to the revolutions of the two planets. Components 4 and 5 have periods of 47,000 and 383,000 years, respectively, which fall neatly into the two slowest regimes of the orbital elements. Physically, these oscillations seem to relate to the eccentricity cycles of the planets, whose periods have been estimated from numerical models to be 45,900 years (Saturn) and 305,000 years (Jupiter) \citep{laskar88}. The discrepancy between these reported values and those obtained from our simulation likely results from the tertiary effects of other planets and moons in the solar system (which were omitted from our model).

The only DMD component that is not closely matched to any of the orbital elements is the third, with a period of 107 years. This may represent some minor resonance phenomenon of the planetary revolution (it is almost exactly 9 Jovian years), but its specific origin is not clear. In any case it is not a dominant effect; this component has the smallest amplitude of all those identified and it could be omitted entirely without dramatically affecting the full reconstruction.

\section{\label{sec:forecasting}Multiscale Forecasting}
\begin{figure}[htb!]
\centering
\includegraphics[width=0.9\columnwidth]{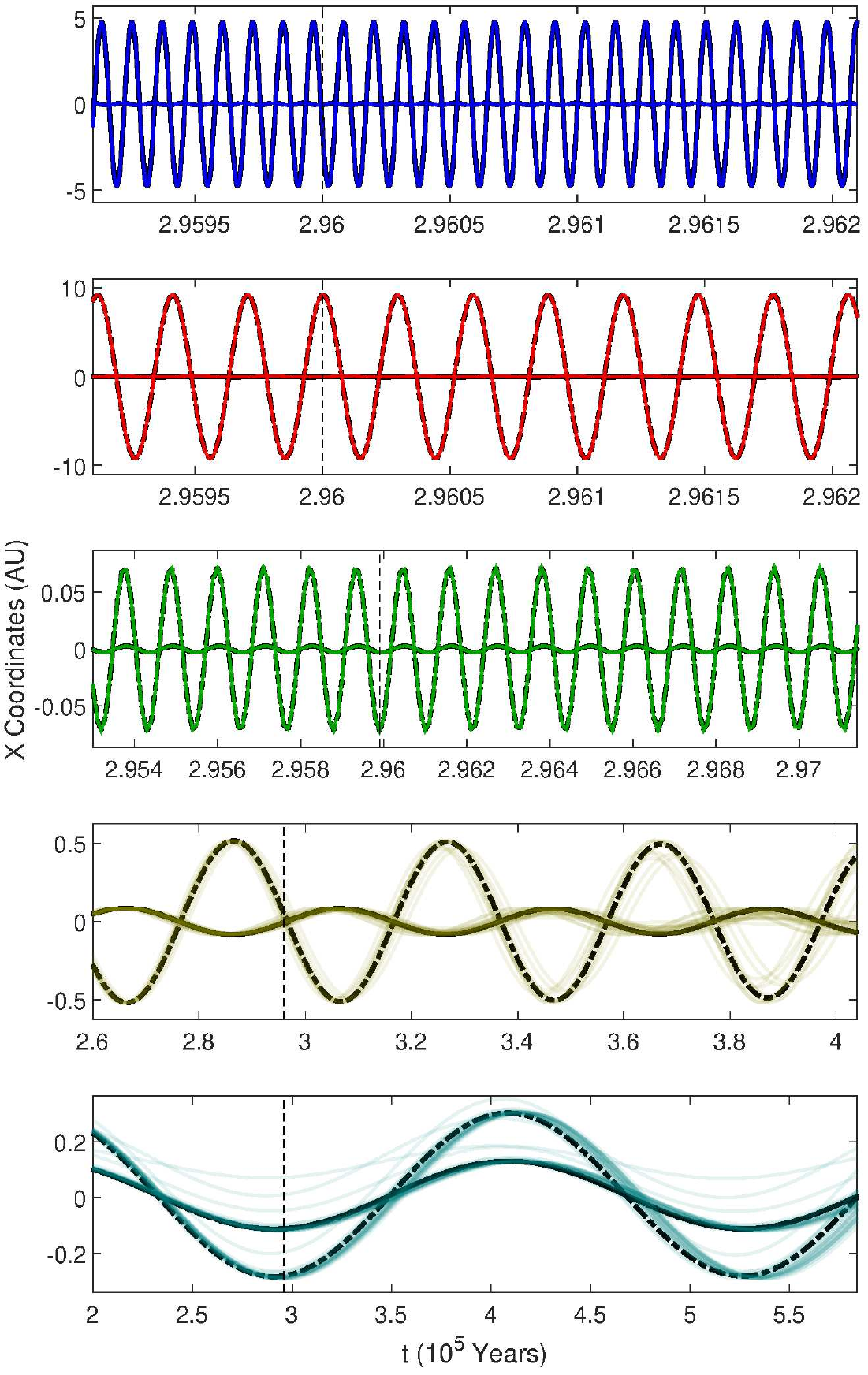}
\caption{Ensemble component-wise forecasts from the sliding-window DMD model on the three-body planetary system. The dotted vertical lines represent the ``current time;'' everything to their right is the forecasted trajectory. At each scale, the unperturbed DMD forecast (black) is overlaid by an ensemble of predictions generated by sampling $b_j^k$, $\boldsymbol{\phi}_j^k$, and $\omega_j^k$ according to their statistics in the preceding windowed iterations (constraining all $\boldsymbol{\phi}_j^k$ to maintain unit length). Note that these trajectories are all nearly identical in the fast-scale components, where the variance of DMD parameters is minimal}
\label{fig:ensemble_forecast}
\end{figure}

Finally, we present a brief example of how the sliding-window DMD approach might be put to practical use. While the decomposition process can be somewhat costly in computational overhead, the execution of the resulting dynamical model is quite efficient (it is simply a closed-form sum of complex exponentials which can be evaluated at arbitrary $t$). The cost-benefit assessment therefore favors applications in which many calls are made to the predictive model. It is thus quite a natural fit to explore the use of this technique as a precursor to Ensemble Kalman Filtering (EnKF). EnKF is a well-established data assimilation technique that integrates measurement data with an ensemble of modeled forecasts \citep{evensen03}. For many models, this ensemble must be built by stochastically perturbing the parameters of some governing differential equation and then numerically integrating out to the target time. With a DMD-based model, however, no integration is necessary. Perturbations can be applied directly to the model parameters $b_j^k$, $\boldsymbol{\phi}_j^k$, and $\omega_j^k$ according to distributions obtained from the spread of those values over all windowed iterations. A very large ensemble could therefore be built quite efficiently, which would be of particular use for online EnKF. A sample of such an ensemble is plotted component-wise in Fig. \ref{fig:ensemble_forecast}.

\section{\label{sec:discussion}Discussion}

We have developed a data-driven method for separating complex, multiscale systems into their constituent time-scale components using a recursive implementation of {\em dynamic mode decomposition} (DMD). The method provides a robust mathematical architecture for regressing to a hierarchy of linear models approximating the nonlinear dynamics at different temporal scales. It even applies to multiscale dynamics produced by coupled, strongly nonlinear oscillators. For integrable systems, for instance, it can extract the constant frequencies of nonlinear oscillations. For nearly integrable systems, these frequencies may no longer be constant, but they change slowly and such variations can be captured by the sliding windows. If there are fast chaotic dynamics, however, then it has no theoretical guarantee to work.
In addition to providing diagnostic information on the frequency content of a signal, our method produces 1) faithful reconstructions of each constituent component with minimal cross-pollution between them, 2) closed-form expressions for these reconstructions which can be used for low-cost forecasting at any time scale, and 3) statistics on the parameters of windowed DMD models, whose distributions can be sampled for stochastic ensemble forecasting. 

\subsection{Connection to Koopman Theory}
The underlying DMD algorithm exploited has a well-documented relationship to Koopman theory, so we briefly comment on how this applies to our technique. The Koopman operator is a linear operator in some measure space which is fully represents the nonlinear dynamics in the original state space of some system. It is typically infinite-dimensional, but can sometimes be well-approximated in finite dimensions. DMD is one of a number of methods which accomplish this: the matrix $A$ which steps the data forward in time plays the same role in $r$ dimensions that the full Koopman operator would in infinite dimensions.

The sliding-window approach presented in this paper generates a new approximator to the Koopman operator for each DMD iteration. As such, the scale discovery protocol can be viewed as an ensemble approach to building a statistical distribution for the Koopman eigenvalue spectrum and identifying peaks that correspond to discrete time scales present in the original data. While the eigenvectors of $A$ vary from one window to the next, and so cannot be interpreted as global Koopman eigenfunctions, the spectral distribution of eigenvalues could serve as a valuable starting point for an algorithm seeking these functions.

\subsection{Utility and Applications}
The recent ascendance of machine learning techniques for analyzing complex systems, along with advances in hardware to support these techniques, has dramatically overhauled engineering approaches for diagnostics and control of such systems. These methods of course require high quality data sets, but also often rely heavily on an interstitial preprocessing step. For time series data with highly disparate time scale content, scale separation is an integral preprocessing procedure for many tasks. Modeling, forecasting, and control of discretely-multiscale systems are much more effective when the scale components can be treated separately. This is particularly true in the common case where the governing dynamics of these components are only weakly coupled to one another: modeling them independently can produce excellent approximations to the true dynamics at a fraction of the computational cost.

With this in mind, the method outlined in this paper is presented as a possible precursor to any data-driven application seeking to exploit a system's multiscale properties. While its output is not entirely dissimilar from well-established multiresolution analysis methods, its differences from other approaches render it particularly well-suited to this role. Its synthesis of spatial and temporal coherencies in the data integrate well into dynamics-focused applications; it can more robustly separate components even when one briefly encroaches on the other's characteristic time scale. Clean separation on this basis is crucial for scale-separated model discovery. Furthermore, it generates a closed-form parametric model for time-local dynamics, which opens up possibilities for forecasting explored in Section \ref{sec:forecasting}. The example presented there is fairly rudimentary, but a more nuanced approach might prove a useful forecasting tool in and of itself. One possible approach is a two-step algorithm which first predicts time evolution of DMD eigenvectors and then builds a full data prediction from those results.

%

\begin{acknowledgments}
JNK acknowledges support from the Air Force Office of Scientific Research (AFOSR) grant FA9550-17-1-0329. MT acknowledges support from NSF DMS-1521667 and DMS-1847802.
\end{acknowledgments}

\bibliography{multires_decompositions_paper_v2}

\end{document}